\documentstyle[11pt,newpasp,twoside,epsf,epsfig]{article}

\def\edcomment#1{\iffalse\marginpar{\raggedright\sl#1\/}\else\relax\fi}
\reversemarginpar

\begin{document}
\title{Interferometric Detection of Planets/Gaps in Protoplanetary Disks}

\author{S. Wolf}
\affil{
California Institute of Technology - JPL/IPAC, 
1200 E California Blvd, Mail code 220-6,
Pasadena, CA, 91125, USA
}

\author{F. Gueth}
\affil{Institut de Radio Astronomie Millim\'etrique, 300 
  rue de la Piscine, 38406 Saint Martin d'H\`eres, France}

\author{Th. Henning}
\affil{Max-Planck-Institut f\"ur Astronomie, K\"onigstuhl 17, 69117 Heidelberg, Germany}

\author{W. Kley}
\affil{Universit\"at T\"ubingen, Inst.\ f\"ur Astronomie und Astrophysik,
Abt. Computational Physics, Auf der Morgenstelle 10,
D-72076 T\"ubingen, Germany}

\begin{abstract}
We investigate the possibility to find evidence for planets in circumstellar
disks by infrared and submillimeter interferometry. 
Hydrodynamical simulations of a circumstellar disk around a solar-type star
with an embedded planet of 1 Jupiter mass are presented.
On the basis of 3D radiative transfer simulations, images of this system are calculated.
These intensity maps provide the basis for the simulation of the interferometers VLTI
(equipped with the mid-infrared instrument MIDI) and ALMA.
While ALMA will provide the necessary basis for a direct gap and therefore 
indirect planet detection, 
MIDI/VLTI will provide the possibility to distinguish between disks with or without accretion 
on the central star on the basis of visibility measurements.
\end{abstract}

\section{Introduction}

Hydrodynamical simulations concerning the evolution of protoplanets in protoplanetary disks
have shown that giant protoplanets may open a gap and cause spiral
density waves in the disk (see, e.g., Kley~1999, Kley et al.~2001, D'Angelo et al.~2002).
Depending on the hydrodynamical properties of the planet and the disk,
the gap may extend up to several AU in width.
We investigate the possibility to find such a gap as an indicator 
for the presence of a protoplanet with present-day or near-future techniques.
For this reason, we use hydrodynamical simulations of a protoplanetary disk 
with an embedded planet and compute the expected brightness distributions.
We show that ALMA will provide the necessary basis to detect gaps 
in circumstellar disks in the millimeter/submillimeter wavelength range.

\section{The disk model}\label{model}

\begin{figure*} 
  \plotone{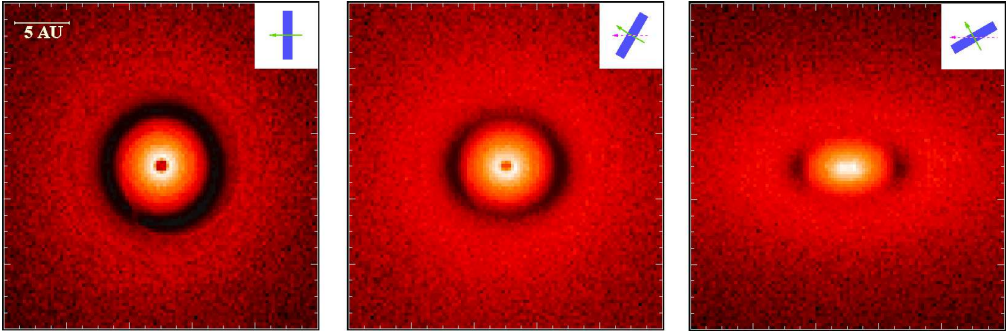}
  \caption{
  Images of the inner region of the circumstellar disk with an embedded planet
  of 1 Jupiter mass at a wavelength of $\lambda$ = 700\,$\mu$m
  and inclinations $i = 0^{\rm o}$ (face-on), $30^{\rm o}$, and $60^{\rm o}$.
  {\sl [taken from Wolf et al.~2002]}
  }
  \label{f1}
\end{figure*}
The density structure of the protoplanetary disk results from hydrodynamical
simulations in which the disk is assumed to be flat and non-self-gravitating.
The mutual gravitational interaction between the planet and the central star,
and the gravitational torques of the disk acting on planet and star
are included. The 3D density structure is Gaussian in the vertical
direction where for the scale height H(r) we assume a constant ratio H/r=0.05. 
The mass of the star is assumed to be 1 ${\rm M}_{\sun}$.
The planet with a mass of 1 Jupiter mass is located at a distance of 5.2\,AU from the star.
The disk has a mass of $0.05\,{\rm M}_{\sun}$ and a diameter of 104\,AU. 
The structure of the spirals and the gap reaches an equilibrium after
about 150 orbits ($\approx 1800\,{\rm yrs}$).

The dust reemission is simulated with the Monte-Carlo radiative transfer code MC3D
(Wolf, Henning, \& Stecklum~1999; Wolf \& Henning~2000; Wolf~2002).
The radiative transfer is simulated self-consistently, taking into account both
the initial temperature of the dust due to viscous heating
and the additional energy input of the central star
(effective temperature $T_{\rm eff}=5500$\,K, luminosity $L = 1\,{\rm L}_{\sun}$).
The dust grains are assumed to consist of ``astronomical'' silicates 
(optical data from Draine \& Lee 1984; radius 0.12\,${\rm \mu}$m;
 dust-to-gas mass ratio = 1:100).
In Figure~1, images of the inner region (diameter 28\,AU) 
of the disk, seen under different inclination angles $i$ at a wavelength of
$\lambda$=700\,$\mu$m, are shown.

\section{Simulations of observations with MIDI (VLTI) and ALMA}\label{simmial}

\begin{figure}[t]
  \plotone{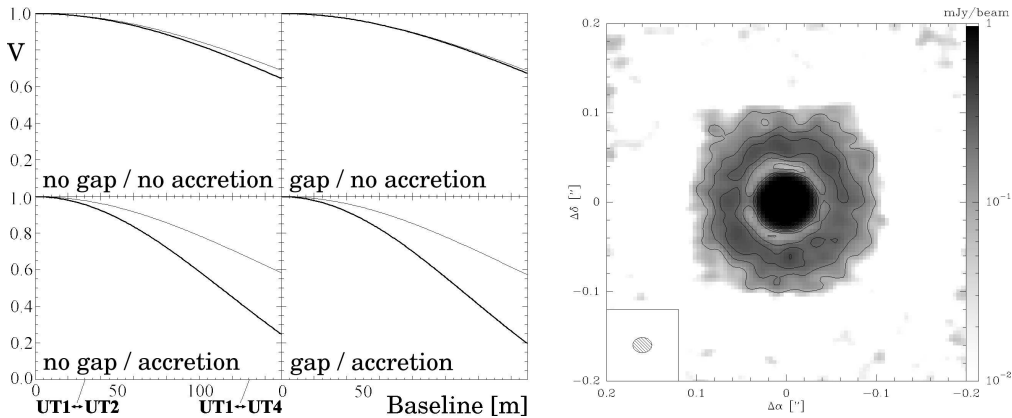}
  \caption{
    {\bf Left:}
    Normalized visibilities at $\lambda$=10\,$\mu$m for a disk with an inclination
    of $i=60^{\rm o}$ (assumed distance: 140\,pc) for a disk with/without a gap and
    with/without accretion.    
    The visibilities marked by the thick and the thin line
    are oriented perpendicular to each other in the u-v plane 
    (parallel to the major/minor axis of the ellipse resulting from 
    the projection of the disk onto the plane of the sky).\\
    {\bf Right:}
    Reconstructed image of the disk seen face-on resulting 
    from a simulation of ALMA (assumed distance: 140\,pc).
    The gap at an angular distance of
    37\,mas (5.2\,AU) from the star is clearly visible.  
    Wavelength: 700\,$\mu$m (428\,GHz);
    bandwidth: 8\,GHz;
    total integration time: 4\,h;
    system temperature: 500\,K;
    phase noise: 30$^{\rm o}$;
    max.\ baseline: 10\,km. {\sl [taken from Wolf et al.~2002]}
  }
  \label{f2}
\end{figure}

The goal of the following simulations is to check whether MIDI/VLTI or ALMA
can be used to detect gaps discussed in the previous sections.

{\bf MIDI:}
This {\bf MID}-infrared two-beam {\bf I}nterferometric instrument is designed for
operation at the Very Large Telescope Interferometer (VLTI) at ESO (Leinert et al.~2000).
In Figure~2, simulated visibility curves at $\lambda$=10\,$\mu$m are shown for 
different disk scenarios. Comparing the results for a disk with/without a gap
we find that the visibilities at a given baseline differ by less than 5\%
since the innermost region -- which dominates the 10\,$\mu$m flux -- 
is only negligibly affected by the presence of the planet at a distance of 5.2\,AU.
Taking into account uncertainties of ``real'' measurements
a distinction between different disk models (with/without gap)
based on significant differences between the visibility profiles
is not possible even by a beam combination of the most distant telescopes of the VLTI.
However, MIDI will be able to trace the presence of accretion onto the central star.
The different density profiles of the innermost region of the disk 
cause a steeper visibility profile
compared to the visibilities simulated for a disk without accretion.
Thus, one could clearly distinguish a disk with accretion from a disk without.

{\bf ALMA:}
The {\bf A}tacama {\bf L}arge {\bf M}illimeter {\bf A}rray
is a planned array of 64 12m-antennas with an aspired maximum baseline of 12 to 14\,km.
It will cover the submillimeter/millimeter wavelength range ($\lambda$=0.3-10\,mm).
Due to the large number of receivers and the resulting broad distribution
of baselines, a sufficient u-v plane coverage can be achieved after
a few hours of observation and image reconstruction will be possible.
In contrast to the first instruments at the VLTI, the visibility
phase can be easily measured in the millimeter wavelength range.
In Figure~2, a reconstructed image of the disk based 
on a simulation of ALMA observations is shown.
Even under consideration of the thermal noise caused by a system temperature
of $T_{\rm sys}$=500\,K, the gap is clearly visible.

\section{Conclusions} \label{concl}

We studied the possibility of the indirect planet detection by observation
of the resulting gap in a protoplanetary disk.
Based on hydrodynamical simulations and subsequent 3D continuum radiative transfer 
calculations we generated images of a circumstellar disk with an embedded 
Jupiter mass planet surrounding a solar-type star.

We found that the gap can be seen very clearly in the simulated
images but an extremely high angular resolution is required ($\approx$\,10\,mas).
Furthermore, because of the extreme brightness contrast in the innermost region 
of the disk in the near to mid-infrared, the gap can hardly be detected 
in this wavelength range with imaging/interferometric observations.
However, we found that it will be possible to distinguish between a disk with or 
without accretion onto the central star with MIDI.
In contrast to this, the (sub)millimeter interferometer ALMA 
will provide the basis for the reconstruction of an image of a gap.
Thus, the search for massive protoplanets in circumstellar disks
can be based on the indication of a gap.

\acknowledgments

\hspace*{1cm}  \\ This research was supported by the DFG grant Kl~650/1-1.


\end{document}